# Real-Time Wave Mitigation for Water-Air OWC Systems Via Beam Tracking

Yujie Di, Yingjie Shao, *Student Member, IEEE,* and Lian-Kuan Chen, *Senior Member, IEEE*

*Abstract*—In a water-air optical wireless communication (OWC) channel, dynamic ocean waves may significantly deflect the light beam from its original direction, thus deteriorating the communication performance. In this letter, a beam tracking system to mitigate wave-induced communication degradation is experimentally demonstrated. By employing the beam tracking on the PAM6 system, a maximum of 486% improvement (from 140 Mb/s to 820 Mb/s) in throughput is realized at a 1.2-m air distance for an average wave slope changing rate of 0.34 rad/s. For PAM4 signals, the packet loss rate reduces significantly from 75% to 11%, and a maximum throughput of 1.25 Gb/s is achieved.

## I. INTRODUCTION

THE underwater exploration for ocean resource excavation and marine ecological system monitoring has intensified, leading to the proliferation of underwater vehicles and sensors [1]. The massive undersea information can be transmitted to an aerial vehicle for data collection or for further relaying the data back to a land station; thus, a high-speed and robust water-air communication link is needed. Optical wireless communication (OWC) systems possess competitive advantages of flexibility, much higher bandwidth, and relatively low attenuation in contrast to wireline, acoustic, and radio frequency (RF) communications, respectively [1]. However, the complex and volatile water-air communication channels resulted from ever-changing surface waves severely degrade the performance of the water-air optical wireless link. The dynamic water wave may deflect the light beam from its original direction and result in the power fluctuation at the receiver. Thus, mitigation of wave-induced impairments is essential to achieve a robust water-air OWC system. An interesting approach that exploited the wave's temporal and spatial characteristics for adaptive bit-loading, achieving a 96.2% increase in throughput, had been proposed [3]. In [4], the beam coverage area was expanded to accommodate the wave-induced beam deflection. Similarly, a diffuse-line-of-sight communication with a data rate over tenths of Mbps was demonstrated over a 0.3-m water link and a 0.6-m air link through a wavy water surface [5]. However, a large beam coverage area implies low power density, unsuitable for long-distance transmission. An effective solution to alleviate the wave impairment without sacrificing power budget is via beam tracking. Laser beam tracking systems had been demonstrated for the applications of inter-satellite and ground-satellite optical link alignment in the 1970s [6]. In 1994, a bidirectional optical link between satellite and ground station was realized, achieving a data rate of 1.024 Mbps. A beam tracking system was implemented to mitigate the power fluctuation, which was mainly comprised of coarse tracking based on a charge-coupled device (CCD) and a fine tracking based on a quadrant detector (QD) [7]. Later, the first inter-satellite OWC link at 50 Mbps was successfully demonstrated. The direction towards the terminal on the other satellite was obtained by detecting the incoming light on CCD [8].

Beam tracking has also been investigated for indoor OWC to direct a narrow collimated beam from the transmitter to the mobile receiver. A widely used method is using a light-emitting diode (LED) array as a tag for recognition and a camera or image sensor for target positioning. A tracking system was designed and implemented for a 50-Gb/s indoor link with a localization accuracy of 0.05° [9]. However, the limited frame rate of the camera restricts the tracking speed. In addition, picture processing for the target recognition increases the system complexity and thus extends tracking response time further. Utilizing a positioning sensitive detector (PSD) or QD is a favorable alternative as it can output position information within 100 μs, but sacrificing the beam tracking area. A 3-D tracking scheme with a reflected beam as a feedback signal was implemented for real-time fast-motion sensing [10]. A PSD was placed on the transmitter side to capture the reflected light from the mobile receiver attached with a retroreflector. Thus, the motion direction of the receiver could be analyzed, and a specific angle was deduced for beam steering to maintain the optical link connection. A bidirectional free-space optical link was established with a similar scheme, capable of automatic connection and keeping track of the moving node [11].

The beam tracking schemes for wave mitigation have not been investigated. The wave-induced dynamic change of light spot location at the receiver plane is rapid and random, requiring higher-speed and more accurate tracking. In [12], we demonstrated the first beam tracking system for water-air OWC systems, and a 50-Mb/s stable optical wireless link was realized

This work was supported in part by HKSAR UGC/RGC grants (GRF 14207220) (*Corresponding author:Yingjie Shao*)

The authors are with the Department of Information Engineering, The Chinese University of Hong Kong, Shatin, N. T., Hong Kong SAR (e-mails: dy019@ie.cuhk.edu.hk, sy017@ie.cuhk.edu.hk, lkchen@ie.cuhk.edu.hk).

under the wavy condition. A low-cost 3×3 photodiode (PD) array was used for light spot positioning to provide a larger detection area, yet compromising its location accuracy.

In this letter, the tracking system is systematically analyzed to optimize the control algorithm. A more detailed study on the effects of wave conditions, data rates, modulation formats, and air-path heights is investigated. Compared with [12], significant improvement in communication performance is achieved for all cases. The results show that by employing the proposed tracking system, the throughput drastically increased by 486% at 600 Mbaud for PAM6 signals, and a maximum throughput of 1.25 Gbit/s at 800 Mbaud is achieved for PAM4.

## II. PRINCIPLE AND EXPERIMENTAL SETUP

### A. Wave Characterization

When a light beam propagates through the water-air interface, the light spot position on the receiver varies with the change of the wave slope, resulting in a time-varying offset from the detector, as shown in Fig. 1(a). We assume a vertical incident light, as denoted by the blue line. At time $t$, the wave slope $k(t)$ at the intersection point is calculated by the derivation of time-varying wave equation $f(t, \vec{x})$. Thus, the wave slope $k(t)$ and the corresponding wave slope angle $\gamma$ can be expressed as

$$k(t) = df(t, \vec{x})/d\vec{x} \quad \text{and} \quad \gamma(t) = \arctan(k(t)). \quad (1)$$

When light passes through the water-air interface, the light refraction angle $\beta$ can be calculated by the incidence angle $\alpha$ using Snell's law

$$n_{water} \cdot \sin\alpha(t) = n_{air} \cdot \sin\beta(t), \quad (2)$$

where $\alpha(t) = \gamma(t)$. Then, the light spot displacement $d$ induced by refraction can be derived from $\alpha$, $\beta$ and the height $h$ from the water surface to the receiver:

$$d(t) = h \cdot \tan\left(\beta(t) - \alpha(t)\right) \quad (3)$$

Since $\alpha$ and $\beta$ only depend on the wave slope angle $\gamma$, the offset $d$ can be expressed as a function of $\gamma$ for a constant $h$. The required tracking speed depends on the light spot drift velocity, $d'(t)$, where $()'$ denotes the differential operator. To capture the key wave characteristic that affects the tracking, we use the time average of the wave slope changing rate (ASCR), which is $\text{ASCR} \triangleq \overline{|\gamma'(t)|}$. We will study the dependence of tracking performance with respect to the ASCR in the following experiment.

### B. Tracking Principles

The schematic of the beam tracking system based on microelectromechanical systems (MEMS) mirror is illustrated in Fig. 2(a). A light beam is emitted from a laser diode (LD) and reflected by a two-axis scanning MEMS mirror that is placed at 45 degrees to the horizontal direction and can be tilted at a specific angle in both x- and y-axis directions. For a tranquil water surface, the light beam will pass through the water-air surface with no deflection. After being divided into two directions by a beam splitter (BS), the two light beams will hit the center of the receiver photodiode (PD) and the corner cube retroreflector (CCR), respectively. The CCR comprises three orthogonal reflective surfaces to reflect light antiparallelly to the incident beam, regardless of the changes in light beam position or incident angle. The beam reflected by the CCR will retrace the incident light path and reach the center PD in a 3×3 PD array at the transmitter side.

When the light beam passes through the wavy water surface, it will be refracted, resulting in a light spot offset at the receiver and the CCR. After being reflected by the CCR, the light is antiparallel to the incident light with some spacing. Thus, the beam will fall on a point away from the center of the PD array at the transmitter side, and the offset serves as a feedback signal for a microcontroller unit (MCU) to control the MEMS mirror tilting angle for correction.

To ensure that the light spot will fall on the center of CCR and PD array simultaneously, the PD array and the MEMS mirror are placed at an equal distance from the BS such that $\overline{AB} = \overline{BD}$, as shown in Fig. 2(c). The proof is as follows. A blue dot line is drawn perpendicularly to the BS. $\theta_1$ equals to $\theta_2$ as the beam is reflected by the BS. As $\theta_3 = \theta_1 = \theta_2$, it can be deduced that $\angle ACB = \angle DCB$. In addition, $\angle ABC = \angle DBC$ as BS is placed at 45°. In addition, since BC is shared by the two triangles, $\Delta ABC$ and $\Delta DBC$, the two are congruent. It implies the center of PD array should be collocated with point D. Similarly, two congruent triangles can be found at the receiver side if the distances from BS2 to CCR and to receiver PD are the same. The light beam that reaches the CCR center will also reach the centers of the PD array and the receiver PD simultaneously.

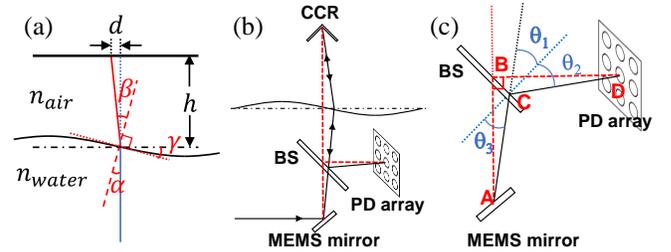

Fig. 1. (a) The beam deflection and light spot offset induced by wave, (b) the MEMS mirror correction adjustment to recenter the beam at the CCR and PD array, and (c) the enlarged view of the transmitter side of (b).

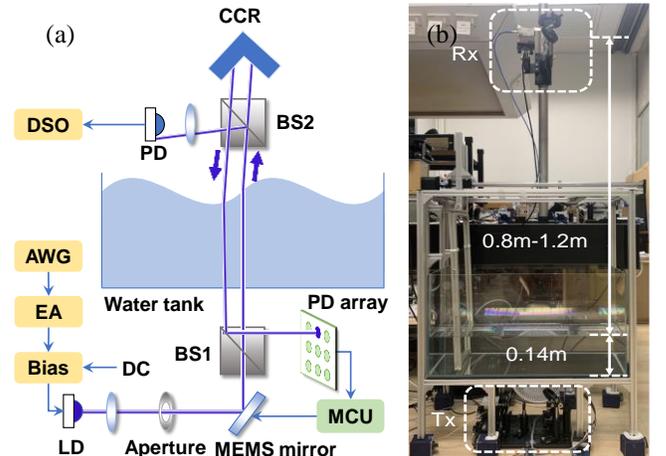

Fig. 2. (a) The proposed tracking scheme and (b) the experimental setup. Tx: transmitter; Rx: receiver.

## C. Tracking Algorithm

After the analog to digital converter (ADC) samples the light intensity captured by the PD array, the tracking process will be triggered if the center PD's light intensity is weaker than a threshold $a$, indicating that the light spot has deviated from the center. A higher threshold means the tracking system is more sensitive to the light spot offset, and the tracking process will be activated earlier. However, it may lead to unintended MEMS mirror adjustment caused by the ambient light fluctuation. Once the tracking process is triggered, the beam offset information can be deduced by locating the PD with the maximum intensity. Suppose the maximum light intensity is larger than a threshold $b$. Then, the MEMS mirror will keep tilting at a specific angle until the light beam is back to the centers of the PD array and the receiver PD, simultaneously. Otherwise, the light spot is assumed lost if the maximum light intensity is less than b for five consecutive measurements. The MEMS mirror will return to the initial point and wait until the PD array recaptures the light spot when the wave slope is back to near 0. Both $a$ and $b$ should be optimized based on the intensity of the light spot and ambient light.

## D. Experimental Setup

The experimental setup of the water-air OWC system with beam tracking is shown in Fig. 2(a) and (b). A water tank with a size of 60 cm×30 cm×40 cm (length ×width ×depth) is filled with tap water of 14-cm depth. It is worth mentioning that our studies focus on the water-air wavy interface and its effect on the air path; thus, the water transmission distance is not the primary consideration in this paper. The wave is generated by a periodically moving plate, controlled by a programmable controller, to stir the water. Hence, by changing the speed of plate movement, waves with different ASCRs can be generated. A 60-second video is recorded to capture the wave-induced light spot displacement at a frame rate of 240 frame/s, by which the wave slope can be calculated for each frame. The instantaneous slope changing rate is then approximated by dividing the wave slope angle difference between two adjacent frames by one frame period. Then ASCR can be derived as in section II-A.

The original electrical signals are generated from an arbitrary waveform generator (AWG) and amplified by an electrical amplifier (EA) of 12-dB gain. The amplified signal, combined with a direct current (DC) bias, is applied to a blue LD (Osram PL450B, 80mW). After being concentrated by a lens, the light is reflected by the MEMS mirror (Mirrorcle, A8L2.2-4600AL-TINY48.4-A/TP), with a diameter of 4.6-mm. The MEMS mirror's tilting angle is controlled by a 16-bit DAC, and the maximum angle is around ±5 degrees for both x- and y-axis. After passing through the water-air channel, the light is divided by a BS with a 50:50 splitting ratio. Half of the light is detected by a 1-GHz avalanche photodiode (APD, Hamamatsu, C5658). The other light beam is reflected by a CCR (Thorlabs, HRR201-P0) to the PD array at the transmitter side. The spacing of two adjacent PD on the PD array is 5 mm. The beam size is set around 7 mm, slightly larger than the PD separation distance to ensure at least one PD can detect the light spot.

The detected signal is then recorded by a digital storage oscilloscope (DSO) for further offline digital signal processing. Two hundred packets, each contains 400K symbols, are collected for analysis under different experimental conditions.

## III. EXPERIMENTAL RESULTS AND ANALYSIS

The normalized intensity of the received optical signal and the light spot displacement trace, with and without the beam tracking, are depicted in Fig. 3. It shows a severe power fluctuation and a dispersive light spot trace induced by the water wave at the receiver side. In contrast, a much more stable and more concentrated received optical power is obtained by employing beam tracking. As the wave was moving in the x-direction, a more significant light spot displacement in the x-axis could be observed.

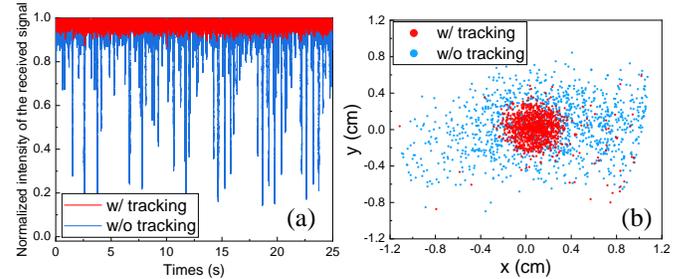

Fig. 3. (a) Power fluctuation and (b) light spot displacement with a 0.8-m air path.

Next, we investigate the average bit error rate (BER) and packet loss rate (PLR) performance for different data rates and air-path distance with OOK and PAM4 signals, as shown in Fig. 4. A packet is considered lost if its BER is below the Soft-Decision Forward Error Correction (SD-FEC) limit ($2×10^{-2}$). The average BER is calculated from all packets, including the lost packets. Clearly, the OWC system with beam tracking outperforms that without tracking for all cases in BER and PLR. The PAM4 signals show a worse BER performance than the OOK signals at the same symbol rate due to its higher requirement of signal-to-noise ratio (SNR). In addition, a longer air-path distance induces a larger beam offset, leading to higher BERs and PLRs. Note that in Fig. 4(a) and (c), the average BER curve without tracking is relatively flat as compared to that with tracking. The reason is that the lost packet's poor BERs dominate the average BER, rather than the increased BERs by a higher data rate. A sharp increase of PLR on 1-Gbaud PAM4 signals is shown in Fig. 4(d), which can be attributed to the limited bandwidth of APD and EA. A maximum PLR reduction, from 75% to 11%, is achieved with 200-Mbaud PAM4 signals for the 1.2-m air path.

We then investigate the BER and PLR performance under different wave ASCR for OOK, PAM4, and PAM6 modulation formats. It is shown that at all ASCR, the measured BER and PLR with beam tracking are lower than that without beam tracking, confirming the effectiveness of the proposed tracking scheme. The larger the ASCR, the higher the BER and PLR. For the system without tracking, the main reason is the larger beam offset. Whereas for the system with tracking, it is mainly due to the limited tracking speed to match the larger ASCR.

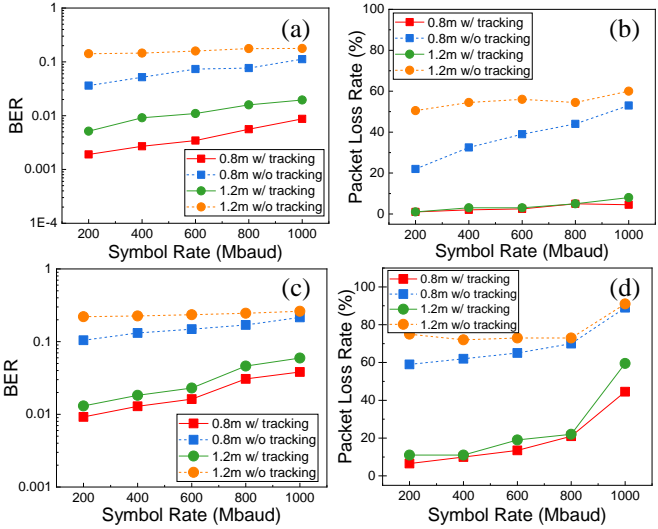

Fig. 4. (a) BER performance and (b) packet loss rate versus symbol rate for OOK signals; (c) BER performance and (d) packet loss rate versus symbol rate for PAM4 signals with the ASCR of 0.34 rad/s.

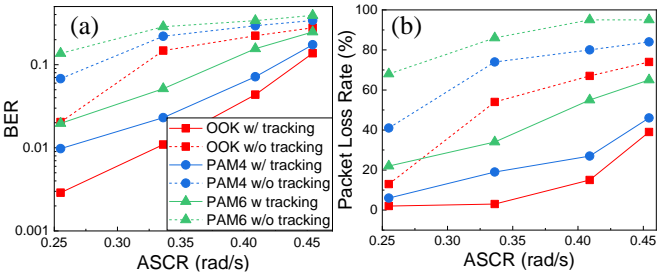

Fig. 5. (a) BER performance and (b) packet loss rate versus ASCR for OOK, PAM4, and PAM6 signals at 600 Mbaud with a 1.2-m air path. The legends of fig. (a) and fig. (b) are the same.

At last, the overall throughputs for different modulation schemes and symbol rates are estimated, as shown in Fig. 6. For an ASCR of 0.34 rad/s, OOK, PAM4, and PAM6 signals achieve a maximum throughput of 920 Mb/s, 1248 Mb/s, and 990 Mb/s, respectively. Hence, PAM4 is the optimal modulation scheme under the current experimental setting. Moreover, a maximum of 130%, 350%, and 486% enhancements in throughput are realized by the proposed beam tracking system for OOK, PAM4, and PAM6 signals, respectively. Note that the PLR of the 1-Gbaud PAM6 signals under static water surface can be as high as 99%.

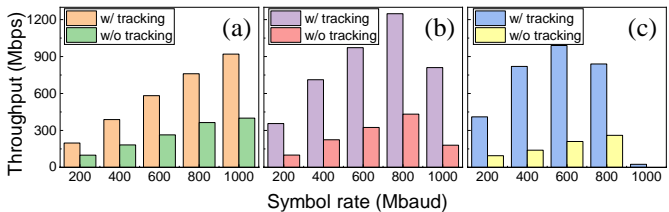

Fig. 6. Throughput versus symbol rate for (a) OOK signals, (b) PAM4 signals, and (c) PAM6 signals with a 1.2-m air path.

## IV. CONCLUSION

In summary, we have proposed and experimentally demonstrated the first beam tracking system to mitigate the wave-induced water-air OWC performance degradation. The effects of symbol rate, modulation format, ASCR, and air-path height are investigated and compared for the OWC system with and without beam tracking. With a 1.2-m air path, the proposed scheme provides a maximum throughput improvement of 486% for PAM6 signals and achieves a maximum throughput of 1.25 Gb/s for PAM4 signals. The wave ASCR determines the required response rate of the tracking system. Albeit the tracking speed limits the performance at a larger ASCR, beam tracking still provides more than 30% reduction in PLR, showing the effectiveness of the proposed scheme in combatting the wave-induced impairment. As the nonlinear relationship between the MEMS mirror tilting angle and the light spot offset, nonlinear control may be applied to attain better beam tracking. This tracking system is also applicable for underwater wireless networks to establish a robust optical link between mobile devices.